\documentclass[draft]{agujournal2019}
\usepackage{url} 
\usepackage{lineno}
\usepackage{amsmath}
\usepackage{amssymb}
\usepackage{booktabs}
\usepackage[inline]{trackchanges} 
\usepackage{soul}
\usepackage{placeins}
\usepackage{float}
\usepackage{multirow}

\journalname{Earth's Future}

\begin{document}

\title{A Comparative Multi-Hazard Risk Assessment of the US High-Voltage Transmission Network}

%
%

\authors{D. Bor\affil{1,2}, E. J. Oughton\affil{1,2}, R. S. Weigel\affil{1,2}, R. Yang\affil{1}, T. Clower\affil{3}, A. Newman\affil{4}, A.R. Valle\affil{4}}

\affiliation{1}{Department of Geography and Geoinformation Sciences, George Mason University, Fairfax, VA}
\affiliation{2}{Space Weather Lab, George Mason University, Fairfax, VA}
\affiliation{3}{Schar School of Policy and Government, George Mason University, Fairfax, VA}
\affiliation{4}{NSF NCAR, Boulder, CO}

\correspondingauthor{E. J. Oughton}{eoughton@gmu.edu}






\begin{keypoints}
    \item We compare ten hazards affecting the US high-voltage transmission grid using one hazard-to-economics framework
    \item The compound freezing rain and wind gust scenario dominates stress-test losses, while tornado leads individual hazards in modeled downstream economic output loss
    \item A 250-year geomagnetic storm produces the same-order output losses as major terrestrial hazard scenarios
\end{keypoints}

\begin{abstract}
Modern economies depend critically on high-voltage power transmission networks. Yet this infrastructure is routinely disrupted by natural hazards ranging from earthquakes and floods to tornadoes and geomagnetic storms. Risk assessments have historically addressed hazards in isolation, leaving no common basis for comparing economic impacts across the full hazard portfolio. This study addresses this gap by developing an integrated framework linking hazard characterization, fragility modeling, and macroeconomic impact propagation. The framework is applied consistently across nine primary hazards and one compound freezing rain and wind gust hazard. Using national hazard datasets and a US high-voltage transmission network of over 13,000 line segments and 10,000 substations, we derive failure probabilities, expected damage, affected population, and downstream economic output losses. Among individual hazards, tropical cyclone wind produces the largest expected daily damage at \$137\,M/day, followed by lightning at \$87\,M/day, earthquake at \$47\,M/day, flood at \$46\,M/day, tornado at \$42\,M/day, and landslide at \$34\,M/day. Downstream economic output losses are largest for tornado at \$4.93\,B/day, followed by flood at \$3.59\,B/day and earthquake at \$3.02\,B/day. A 250-year geomagnetic storm produces \$2.07\,B/day, placing space weather within the range of major terrestrial hazards. The compound freezing rain and wind gust scenario produces the largest stress-test disruption, affecting 237.4\,M people and yielding a modeled downstream output loss of \$85.16\,B/day. These results should be interpreted as first-order bounding estimates, with the compound scenario representing an upper-bound stress test. Overall, the framework establishes a consistent baseline for prioritizing investments in transmission network resilience.
\end{abstract}

\section*{Plain Language Summary}
High-voltage transmission lines and substations are exposed to many natural hazards, including earthquakes, floods, tornadoes, wildfires, lightning, freezing rain, and geomagnetic storms. These hazards have been examined separately in the literature, making it hard for grid planners and policymakers to compare risks across hazards or decide where resilience investments matter most.

We develop a common method for comparing natural-hazard risks to the US high-voltage transmission network. The method integrates national hazard maps, grid asset data, damage models, replacement costs, population exposure, and an economic model. The coupled model estimates how power disruptions affect different sectors of the economy. We find that tropical cyclone wind, lightning, earthquake, flood, tornado, and landslide are the leading individual hazards for physical grid damage. Tornado, flood, and earthquake produce the largest modeled economic losses among the individual terrestrial hazards. A rare geomagnetic storm can cause losses comparable to those from several major terrestrial hazards. A compound freezing rain and wind gust scenario produces the largest disruption. However, this should be treated as an upper-bound stress test, as the hazard data are only available at the state level. These results provide a first-order comparison of power-grid risks across the full hazard portfolio and can support future resilience planning.

\section{Introduction}
\label{sec:multi:intro}

The power transmission network is the backbone of modern economic activity. It supports computing, manufacturing, transport, finance, and the data centers that power the ongoing expansion of artificial intelligence. However, this infrastructure is frequently exposed to natural hazards, including earthquakes, floods, tornadoes, wildfires, and geomagnetic storms. Each hazard poses distinct physical stresses on transmission lines and substations.

The literature has largely addressed these threats in isolation. Seismic fragility of substations, flood exposure of distribution infrastructure, and wind vulnerability of transmission towers have each received sustained attention. For extreme events with return periods exceeding 100 years, however, the economic consequences of transmission failure remain poorly quantified across this hazard portfolio. Moreover, studies evaluating these threats within a common analytical framework remain scarce. Existing assessments are methodologically inconsistent. They employ different intensity measures, fragility formulations, and economic models, making cross-hazard comparison effectively impossible. Thus, infrastructure operators and policymakers allocating resilience budgets across competing threats lack the comparative risk metrics needed to make informed decisions \cite{rokhideh_rethinking_2025}.

Space weather adds a further dimension to this problem. Geomagnetic storms induce quasi-DC currents in long transmission lines that can saturate high-voltage transformers. In extreme cases, they can trigger cascading blackouts. The 1989 Hydro-Qu\'ebec collapse, for instance, left six million people without power for nine hours \cite{pulkkinen_statistics_2008}. Socioeconomic impact assessments for space weather have grown in recent years, spanning the UK \cite{oughton_risk_2019}, the US \cite{worman_chapter_2018}, and New Zealand \cite{mac_manus_implementing_2025}. However, these assessments remain methodologically isolated from terrestrial hazard analyses. This makes it difficult to answer a basic policy question: how does a 100-year geomagnetic storm compare, in economic terms, to a 100-year flood or earthquake affecting the same network?

We address this question by extending the physics-to-economics framework of \citeA{oughton_major_2025} to eight terrestrial natural hazards: earthquake, riverine flood, landslide, wildfire, lightning, tropical cyclone wind, hail, and tornado. The framework couples per-substation geomagnetically induced current (GIC) estimates with transformer-specific fragility functions and input--output economic propagation. The same network representation, fragility database, and economic model are applied consistently across all hazard types. This produces comparable metrics for expected daily damage, affected population, and sectoral output loss. Space weather serves as a comparator within this landscape. This paper addresses the following research questions:

\begin{description}
    \item[RQ1] Which transmission assets and regions face the highest multi-hazard risk, and what failure modes dominate under each hazard type?

    \item[RQ2] How do hazard-induced transmission failures propagate to sectoral economic output loss, and which sectors are most exposed?

    \item[RQ3] How do geomagnetic storm impacts on the US transmission network compare quantitatively to terrestrial natural hazards at equivalent return periods?
\end{description}


\subsection{Hazard-specific fragility and risk assessment}\label{sec:multi:background}
\label{sec:hazard-specific}

The vulnerability of power grid components to individual natural hazards has received substantial attention. The literature remains organized by hazard type, with each community employing distinct intensity measures, fragility formulations, and performance metrics.

Seismic fragility analysis of substations has the longest history. The Hazards United States (HAZUS) methodology \cite{fema_hazus_earthquake_2022} has long provided the most widely used standardized baseline across voltage levels. \citeA{hwang_seismic_1998} developed early analytical fragility curves for equipment and structures at a Memphis substation. This work relied on analytical methods because earthquake damage data were scarce for eastern US facilities. More recent studies have constructed configuration-specific models that account for internal component layout, line-level faults, and short-circuit modes. These studies find that fragility can vary significantly across substation archetypes \cite{ahumada_seismic_2026}. Network-scale frameworks have extended this further to system-level seismic risk and retrofit planning \cite{liang_multi-model_2026}. They have also supported substation resilience quantification through transmission capacity metrics with recovery simulation \cite{liang_seismic_2023}. Distribution systems have attracted comparatively less attention because the literature has historically prioritized transmission infrastructure \cite{liu_quantifying_2021}. 


Wind and hurricane fragility research has focused heavily on lattice transmission towers. Pushover-based approaches have yielded generalized fragility parameters for 154, 345, and 765~kV Korean configurations \cite{woo_proposal_2026}. Probabilistic system-level methods show that angled tower--line systems can be up to 140\% more vulnerable than straight-line configurations because of spatial variability in wind loading \cite{feng_probabilistic_2025}. This coupling between tower and conductor response proves critical at the network scale. \citeA{xue_impact_2020} showed that coupled fragility models reproduced observed damage during Hurricane Harvey on the Texas 2000-bus network, whereas isolated tower models could not. Neglecting the mechanical interaction between tower and line misrepresents how loads are distributed across the system. Component-based methods have further disaggregated failure into member-level buckling and yielding \cite{ma_component-based_2021}. Performance-based limit states spanning five damage levels from intact to collapse have been defined via logistic regression \cite{stewart_hurricane_2022}. Despite this progress, \citeA{hou_review_2022} observes that transmission- and distribution-level models have largely developed in isolation. This leaves a persistent gap between component fragility and system-wide outage prediction.

Flood risk to electrical infrastructure has been assessed at multiple scales. Geographic information system (GIS)-based tools applied to Barcelona showed that drainage improvements can reduce substation failure probability by 50\% and costs by 77\% under climate-change scenarios \cite{sanchez_munoz_gis-based_2021}. \citeA{asaridis_probabilistic_2025} coupled stochastic flood inundation with power flow and socioeconomic models on an Institute of Electrical and Electronics Engineers (IEEE) 14-bus Italian benchmark. \citeA{prieto-miranda_high-resolution_2024} modeled Hurricane Florence's impacts on North Carolina's 662-node grid. Their work showed how localized substation inundation propagates to system-wide congestion. Pre-event protection strategies have been formalized through stochastic resource allocation for the deployment of temporary flood barriers around critical substations \cite{movahednia_power_2022}. Less common sub-hazards include flash-flood impacts on distribution systems \cite{afzal_modeling_2024} and tsunami-bore loading on utility poles, quantified through flume experiments and incremental dynamic analysis \cite{stephens_vulnerability_2023}.

The wildfire--power grid nexus operates bidirectionally. Wildfires damage transmission assets through thermal degradation and conductor flashover. Grid infrastructure can also ignite fires through conductor--vegetation contact. \citeA{wang_predicting_2023} modeled this encroachment probability through spectral analysis of a dynamic conductor displacement, finding it highly sensitive to vegetation clearance and wind intensity. Machine learning approaches trained on Pacific Gas and Electric (PG\&E) data from 2015--2019 achieved area under the curve (AUC) scores of 0.776 for distribution feeder ignitions and 0.824 for transmission wire-down events \cite{yao_predicting_2022}. Spatio-temporal frameworks combining deep neural network ignition prediction with physical spread models have been demonstrated on the IEEE~24-bus system \cite{umunnakwe_datadriven_2022}. \citeA{arab_three_2021} organizes this body of work into prevention, mitigation, and recovery. \citeA{vahedi_wildfire_2025} examines how climate change is compounding wildfire-grid interactions across the western United States.

Coverage of other terrestrial hazards remains markedly uneven. Tornado fragility analysis shows vulnerability even to Enhanced Fujita (EF) scale EF1--EF2 events, with pole aging substantially reducing resistance \cite{braik_risk-based_2019}. Landslide impacts on transmission corridors have been quantified through system performance evaluation on a real network \cite{ghorani_modeling_2021}. Lightning, though a leading cause of transmission trips, is predominantly addressed through insulation engineering. System-level exceptions include Bayesian network frameworks jointly analyzing tripping from multiple hazards \cite{bian_risk_2024} and empirical attribution of outage rates to meteorological thresholds \cite{rimkus_impact_2024}. Hail damage to grid infrastructure remains essentially unquantified.

GICs pose a physically distinct threat through half-cycle saturation and thermal heating of high-voltage transformers. National-scale assessments have quantified GIC exposure for Spain \cite{torta_assessing_2014}, Mexico \cite{caraballo_first_2020}, and China \cite{lin_risk_2020}. Carrington-scale scenarios suggest three-phase currents of 25--150~A. The catastrophic potential of such events was outlined by \citeA{zurbuchen_how_2012}. These events could involve widespread transformer failures that leave millions without power for months. Most recently, \citeA{oughton_major_2025} estimated daily losses of \$1.37~billion with a 95\% confidence interval of \$1.16--1.58~billion for a 100-year US storm affecting 4.1~million people. Losses rise to \$2.09~billion for a 250-year event. Despite these advances, space weather assessments remain methodologically disconnected from terrestrial hazard analyses. This precludes direct cross-hazard comparison at equivalent return periods.

\subsection{Multi-hazard frameworks and resilience assessment}
\label{sec:multi-hazard-frameworks}

While hazard-specific studies have matured considerably, efforts to assess power grid vulnerability across multiple hazards within a common framework remain limited. \citeA{salman_multihazard_2017} developed an early multi-hazard reliability assessment combining hurricane and seismic fragility with topological system performance evaluation. This was later extended to a life-cycle cost framework for mitigation strategies on notional networks in Charleston and New York \cite{salman_probabilistic_2018}. At the statistical-empirical end, \citeA{mukherjee_multi-hazard_2018} trained data-mining models on historical US outage records. The study found that risk depends on hazard type, overhead system extent, land use, and maintenance investment. \citeA{reed_multi-hazard_2016} proposed logit-based system-level fragility functions for simultaneous weather hazards, applied empirically to Hurricanes Isaac and Sandy.

At the component level, multi-hazard fragility research has focused on transmission towers under combined wind and seismic loading. \citeA{jeddi_multihazard_2022} developed typhoon--earthquake collapse fragility surfaces using active learning on a double-circuit tower. Their results show that prior damage substantially increases susceptibility to subsequent events. \citeA{li_lifetime_2022} incorporated wind-induced fatigue into lifetime fragility analysis of a 1000~kV tower, demonstrating significant capacity deterioration over service life. \citeA{guo_multi-hazard_2025} found that single-indicator assessments can be misleadingly sensitive to one hazard under simultaneous wind and earthquake loading. Compound climate hazards have also received attention. \citeA{macheri_assessing_2025} modeled wind--flood dependence through a Gumbel copula on the IEEE~24-bus system, revealing voltage collapse and severe line overloads under joint scenarios. \citeA{jackson_spatio-temporal_2025} used Markov random fields to capture wildfire--flood cascading effects on outage probabilities across California counties. The RIESGOS project \cite{pittore_towards_2020} demonstrated a modular, web-service-based approach to multi-hazard scenario assessment. This included power line cascades for earthquake--tsunami--volcanic sequences in South America.

The resilience assessment literature has developed in parallel. \citeA{panteli_power_2017} coupled weather-dependent fragility curves with optimal power flow and sequential Monte Carlo simulation to assess windstorm impacts on Great Britain's transmission network. Companion work introduced the resilience trapezoid and time-dependent operational and infrastructure resilience metrics \cite{panteli_metrics_2017}. \citeA{espinoza_multi-phase_2016} formalized a four-phase framework covering threat characterization, vulnerability assessment, system reaction, and restoration. The framework evaluates when a stronger, larger, or smarter grid is preferable under windstorms and floods. \citeA{mujjuni_evaluation_2023} demonstrated that internal systemic maloperations can dominate over exogenous hazard intensity in determining network outturn. \citeA{babu_comprehensive_2025} provides a recent comprehensive review introducing complex network theory as a complementary quantification tool for distribution systems.

Several reviews have cataloged fragility functions available for resilience studies. \citeA{serrano-fontova_comprehensive_2023} compiled the first systematic cross-hazard comparison, showing that fragility model choice significantly affects resilience estimates. \citeA{karagiannakis_fragility_2025} reviewed analytical and empirical models across transmission, distribution, and substations, highlighting compound hazard gaps. The Pacific Northwest National Laboratory (PNNL) resource report \cite{kabre_fragility_2022} documented fragility functions for electricity and water infrastructure to support investment-grade resilience valuation. Despite this growing body of work, no study has applied a single consistent framework with a common network representation, fragility database, and economic model across the full portfolio of natural hazards threatening the transmission grid.

\subsection{Economic impact propagation}
\label{sec:economic-impact}

Most fragility and resilience studies end at quantifying physical damage or load shedding. The translation into economic consequences remains largely unaddressed.

Direct damage costs have been incorporated into flood risk tools through repair estimates, business interruption, and non-supplied energy valuation \cite{sanchez_munoz_gis-based_2021, movahednia_power_2022}. They have also been incorporated into tornado hazard analysis through life-cycle cost analysis of utility pole materials \cite{braik_risk-based_2019}. At the system level, \citeA{ouyang_multi-dimensional_2014} coupled hurricane hazard, fragility, and restoration models for Harris County, Texas. The framework was calibrated against Hurricane Ike and estimated annualized losses of up to \$83~million despite system resilience values exceeding 99.7\%.

Indirect losses propagating through supply chains have been addressed almost exclusively within the space weather community. \citeA{oughton_quantifying_2017} showed that direct electricity disruption accounts for only 49\% of total potential losses from transmission failure. The remainder arises from supply chain effects across trading partners. This methodology was extended to the United Kingdom, where a Carrington-scale 1-in-100-year event was estimated to cause gross domestic product (GDP) losses of up to \pounds15.9~billion without forecasting. With enhanced space weather prediction, these losses are reducible to \pounds0.9~billion \cite{oughton_risk_2019}. \citeA{oughton_major_2025} subsequently developed a full physics--engineering--economic framework, estimating daily losses of \$1.37~billion for a 100-year US storm affecting 4.1~million people.

Terrestrial hazard studies that incorporate economic dimensions stop short of macroeconomic input--output modeling. These include customer impact models in flood frameworks \cite{asaridis_probabilistic_2025, prieto-miranda_high-resolution_2024} and state-level outage predictors \cite{mukherjee_multi-hazard_2018}. This methodological asymmetry makes the cross-hazard comparison of economic risk at equivalent return periods infeasible. The present study addresses this gap by extending the framework of \citeA{oughton_major_2025} to eight terrestrial natural hazards. It applies the framework consistently alongside the same network representation and fragility database used for space weather assessment.

\newpage
\section{Methodology}
\label{sec:multi:methods}

\begin{figure}[H]
    \centering
    \includegraphics[width=\textwidth]{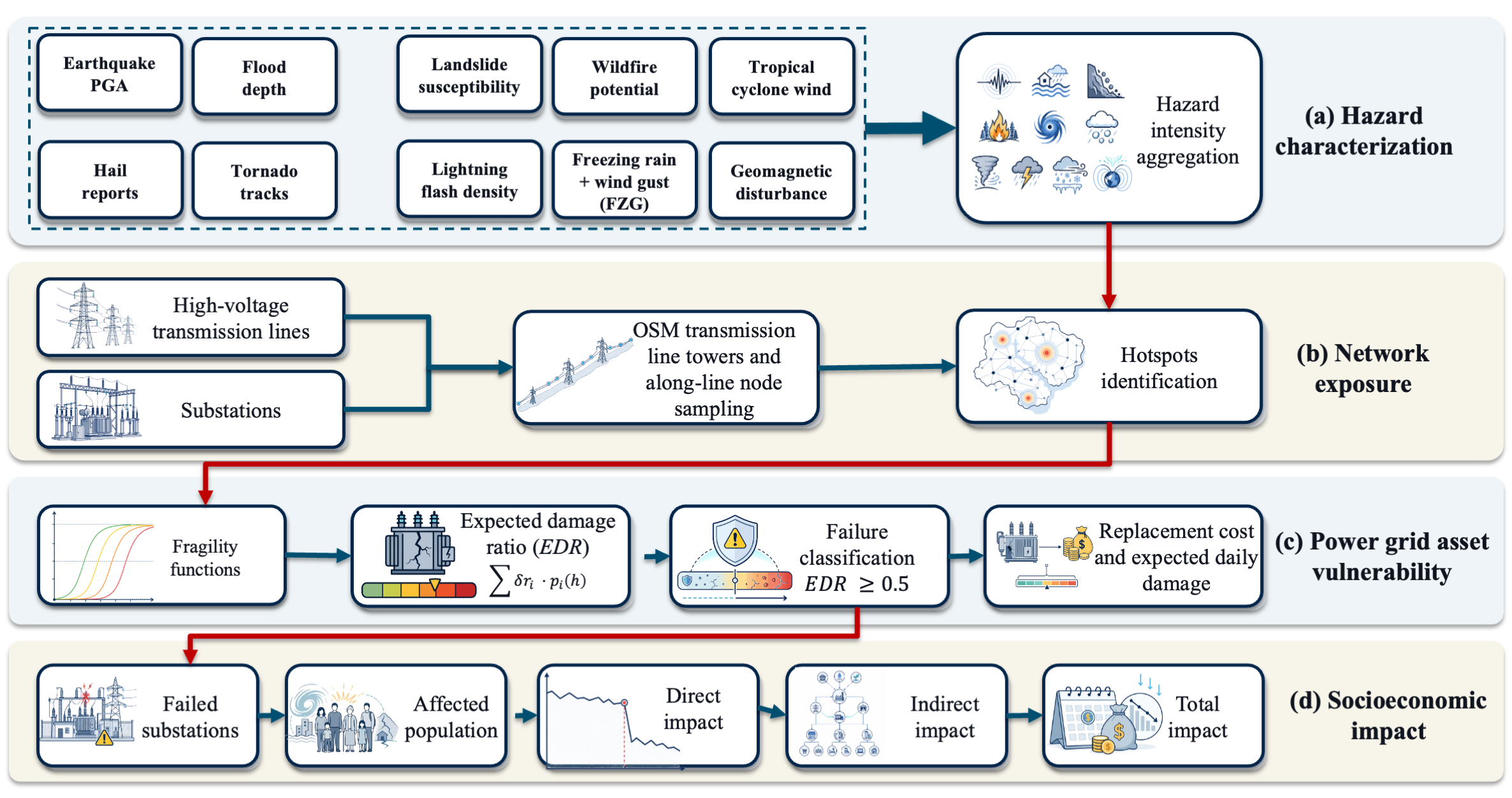}
    \caption{\textbf{Multi-hazard power transmission risk framework.} The workflow links hazard characterization, network exposure sampling, fragility-based damage estimation, failure classification, affected-population estimation, and Leontief input--output economic propagation.}
    \label{fig:method_diagram}
\end{figure}

The methodology integrates hazard characterization, network exposure, fragility-based damage estimation, and downstream economic propagation into a single workflow. It is applied consistently across nine primary hazards and one compound freezing rain and wind gust (FZG) hazard. Each hazard intensity layer is sampled at the same set of network assets. The sampled values are passed through hazard-specific fragility functions to produce expected damage ratios (EDR). The resulting failed-substation set is then propagated through a population-weighted Leontief input--output model to estimate sectoral economic loss. Figure~\ref{fig:method_diagram} summarizes the end-to-end pipeline.

\subsection{Transmission network assets}
\label{sec:multi:assets}

The transmission network is reused from \citeA{oughton_major_2025} and is constructed from two primary data sources. Transmission lines are obtained from the Homeland Infrastructure Foundation-Level Data (HIFLD) database \cite{hifld_transmission_2023}, filtered to operating voltages $\geq$161\,kV. Substations are extracted from OpenStreetMap (OSM) \cite{openstreetmap_openstreetmap_2024} and geospatially linked to transmission lines to derive the network topology. We adopt this network directly to ensure consistency with the geomagnetic hazard pipeline. That pipeline is also built on the same asset base.

\begin{figure}[H]
    \centering
    \includegraphics[width=\textwidth]{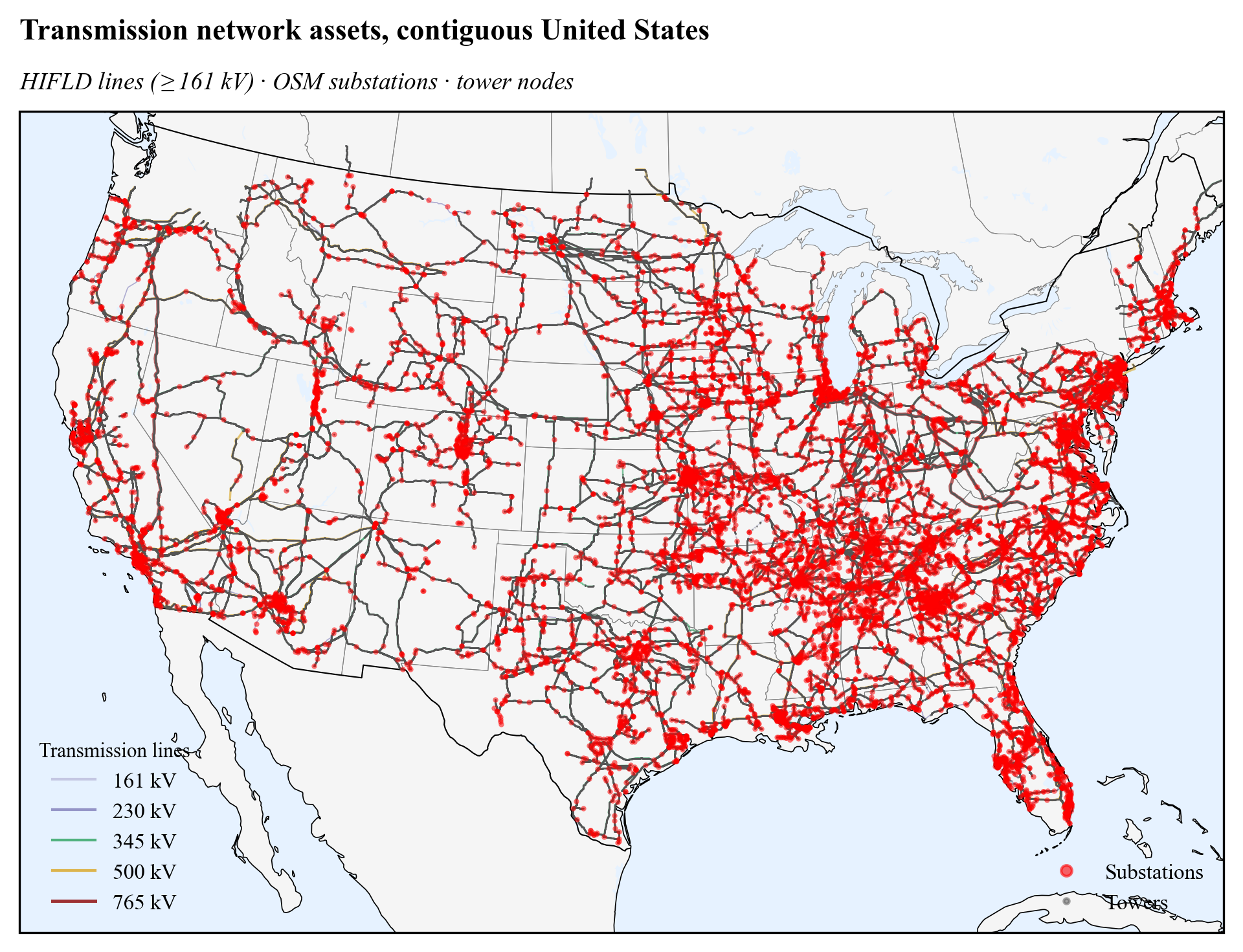}
    \caption{\textbf{Transmission network assets across the contiguous United States.} Lines are colored by voltage class (161--765\,kV). Red points indicate substations. The network is reused from \citeA{oughton_major_2025}, augmented with along-line node sampling for spatially distributed hazard exposure.}
    \label{fig:network_assets}
\end{figure}

For this study, the network is augmented with a denser set of along-line sampling locations. This enables spatially distributed sampling of hazard intensity along transmission corridors. OSM power tower locations are first matched to HIFLD lines via a 250\,m buffer spatial join. Gaps between consecutive towers exceeding 400\,m are then filled with synthetic node points at 200\,m spacing. A 50\,m deduplication threshold is applied to avoid redundancy near existing towers. This produces a hybrid set of real-tower and synthetic-node locations that densely sample each line's geometry. The final network representation used in this study is shown in Figure~\ref{fig:network_assets}.

\subsection{Hazard intensity layers}
\label{sec:multi:hazards}

We characterize nine natural hazards and one compound hazard across the transmission network. All hazard layers are reprojected to EPSG:5070 (Albers Equal Area Conic) to match the asset coordinate reference system. Intensity values are sampled at node points and substation centroids where raster data are available. For gridded tabular sources, nearest-neighbor interpolation is used instead. Node-level values are aggregated to their parent transmission line using mean, maximum, and 95th percentile statistics. Hotspot assets are defined as those exceeding the 95th-percentile exposure threshold for each hazard. The discrete-tier compound hazard, FZG, is the exception. For FZG, hotspots are defined categorically at the catastrophic tier as described in Section~\ref{sec:multi:fzg-method}. Per-hazard exposure maps for all ten hazards are shown in Figure~\ref{fig:multihazard_exposure}.

\subsubsection{Earthquake}

Peak ground acceleration (PGA) at the 475-year return period (2\% probability of exceedance in 50 years) is obtained from the 2023 United States Geological Survey (USGS) National Seismic Hazard Model \cite{mark_d_petersen_data_2024}. The dataset is provided as gridded point data at 0.2$^\circ$ resolution. The reference site condition is $V_{s30} = 760$\,m/s. This corresponds to the National Earthquake Hazards Reduction Program (NEHRP) site class B/C boundary representing firm rock. Because the source data are tabular rather than raster, PGA values are assigned to asset locations via nearest-neighbor interpolation. The PGA, in units of $g$, serves as the intensity measure. The resulting seismic exposure of the network is shown in Figure~\ref{fig:multihazard_exposure}a.

\subsubsection{Flood}

Riverine flood depth at the 100-year return period is obtained from the World Resources Institute (WRI) Aqueduct Global Flood Model \cite{reig_aqueduct_2013} at approximately 1\,km resolution. Inundation depth in meters serves as the intensity measure. The flood exposure of the network is shown in Figure~\ref{fig:multihazard_exposure}b.

\subsubsection{Landslide}

\begin{figure}[H]
    \centering
    \includegraphics[width=\textwidth]{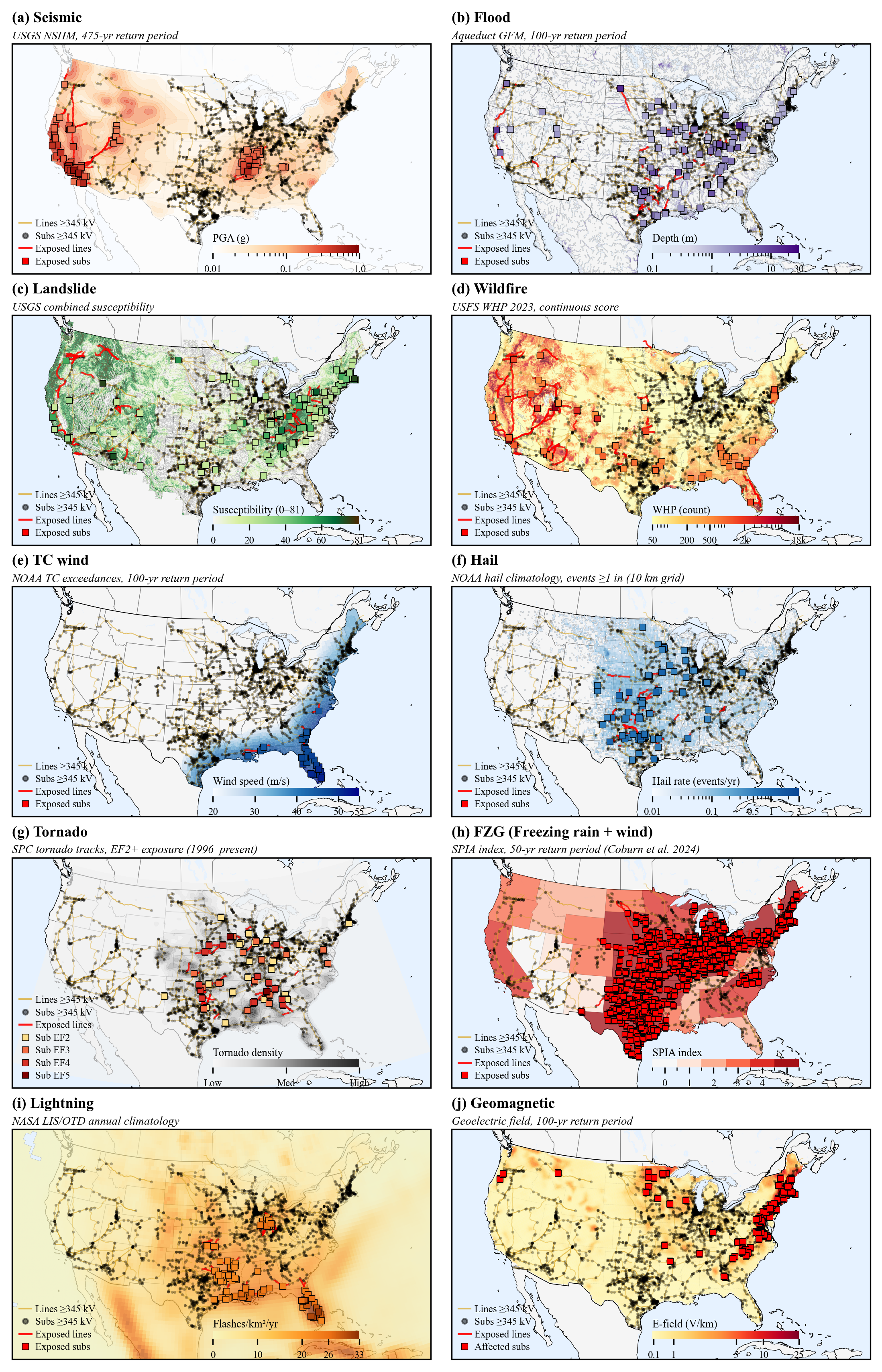}
    \caption{\textbf{Per-hazard exposure of the US high-voltage power network.}}
    \label{fig:multihazard_exposure}
\end{figure}

Landslide susceptibility is obtained from the USGS national assessment \cite{gina_m_belair_slope-relief_2024}, which provides two complementary contiguous United States (CONUS) rasters. The first is a neighborhood count index (n10, range 0--81), representing the number of geomorphic and geologic susceptibility factors within a 10-cell spatial neighborhood. The second is a landslide warning level (lw) derived from precipitation thresholds. The two layers capture distinct aspects of susceptibility: static terrain predisposition (n10) and dynamic triggering potential (lw). A combined susceptibility score is computed as the equal-weight arithmetic mean of the two band values at each sample point. This treats both components as equally informative in the absence of empirical calibration data. The resulting landslide exposure is shown in Figure~\ref{fig:multihazard_exposure}c.

\subsubsection{Wildfire}

Wildfire Hazard Potential (WHP) is obtained from the 2023 US Department of Agriculture (USDA) Forest Service national dataset \cite{dillon_wildfire_2023} at 270\,m resolution. Two bands are provided: a continuous hazard potential index (whp\_cnt, range 0--18,176) and a classified layer (whp\_cls, 1--7). Pixels classified as non-burnable (class 6) or water (class 7) are masked before sampling. The continuous index serves as the intensity measure for fragility analysis. The wildfire exposure of the network is shown in Figure~\ref{fig:multihazard_exposure}d.

\subsubsection{Tropical cyclone wind}

Tropical cyclone (TC) wind speed at the 100-year return period (1\% annual exceedance probability) is obtained from the Synthetic Tropical Cyclone Generation Model (STORM) dataset \cite{bloemendaal_globally_2022}. The data are provided as county-level maximum sustained wind speeds in knots. Wind speeds are converted to m/s and spatially joined to assets via county polygon intersection. Coverage is limited to counties within the tropical cyclone risk zone, which includes 995 of 3,108 CONUS counties. These counties are predominantly along the Gulf Coast and Atlantic seaboard. Wind speed in m/s serves as the intensity measure. The TC wind exposure of the network is shown in Figure~\ref{fig:multihazard_exposure}e.

\subsubsection{Hail}

Hail exposure is derived from the National Oceanic and Atmospheric Administration (NOAA) Storm Prediction Center (SPC) storm reports spanning 1955--2024 \cite{noaa_storm_prediction_center_noaa_2024-1}. The data are filtered to significant hail ($\geq$1.0\,inch diameter, 251,933 events). Reports are provided as point locations using LineString geometries with zero path length. Centroids are extracted, and the resulting point locations are rasterized onto a 10\,km grid to produce two surfaces: annual hail event rate (count per 70 years of record) and maximum observed hail diameter (inches). Both surfaces are sampled at nodes and substations. Maximum hail diameter serves as the intensity measure for fragility analysis. The annual event rate is used as the occurrence frequency, $\lambda$, in loss estimation. The hail exposure of the network is shown in Figure~\ref{fig:multihazard_exposure}f.

\subsubsection{Tornado}

Tornado exposure is derived from 35,104 SPC tornado tracks spanning 1996--2024 \cite{noaa_storm_prediction_center_noaa_2024}, filtered to known magnitudes (EF0--EF5). Each track is buffered by its reported path width. Buffered swaths are spatially joined to node points and substations to identify direct intersections. Per-asset metrics include total hit count, EF2+ hit count, and annual EF2+ strike rate normalized by 29 years of record. EF ratings are converted to estimated wind speeds for fragility analysis, ranging from EF0 = 29\,m/s to EF5 = 89\,m/s. The annual hit rate serves as the occurrence frequency $\lambda$ in the loss estimation. The maximum EF-derived wind speed serves as the intensity measure. The tornado exposure of the network is shown in Figure~\ref{fig:multihazard_exposure}g.

\subsubsection{Lightning}

Annual flash rate density is derived from the NASA Lightning Imaging Sensor / Optical Transient Detector (LIS/OTD) combined satellite climatology \cite{nasa_earth_science_data_systems_lisotd_2024}. The dataset is provided as monthly mean flash rates (flashes/km$^2$/day) on a 0.5$^\circ$ global grid. Monthly values are multiplied by the number of days per month and summed to produce an annual flash rate density (flashes/km$^2$/yr). The annual lightning flash rate serves as the intensity measure. The lightning exposure of the network is shown in Figure~\ref{fig:multihazard_exposure}i.

\subsubsection{Geomagnetic disturbance}

Geomagnetic hazard intensity at the 100-year return period is obtained from the end-to-end physics-engineering coupling of \citeA{oughton_major_2025}. Their methodology derives extreme geoelectric field magnitudes (V/km) across CONUS by combining geomagnetic time series from INTERMAGNET observatories with three-dimensional ground conductivity. These conductivity data come from magnetotelluric measurements collected under the USArray EarthScope campaign \cite{kelbert_em_2011}. These fields are then propagated through a geospatial representation of the bulk high-voltage transmission network. This computes the effective GIC (A/phase) at each transformer under extreme-value return-period scenarios (100-year and 250-year). Per-transformer failure probability is estimated using a lognormal fragility curve anchored at 75 A/phase. This value is the benchmark GIC defined by the North American Electric Reliability Corporation (NERC) standard TPL-007 \cite{nerc_tpl-007-4_2020}. Above this value, transformers must undergo thermal-impact assessment to demonstrate adequate withstand capability. \citeA{oughton_major_2025} adopt it as the median fragility capacity, representing a 50\% probability of transformer maloperation. This fragility structure is consistent with the lognormal framework adopted for terrestrial hazards in this study (Section~\ref{sec:multi:fragility}). We use the per-substation failure probabilities published by \citeA{oughton_major_2025} directly rather than reimplementing the upstream geophysical and reliability pipeline. The hazard intensity measure used in this study is the 100-year return period peak geoelectric field magnitude per substation, shown in Figure~\ref{fig:multihazard_exposure}j. A 250-year scenario from the same source is examined as a tail-risk sensitivity in Section~\ref{sec:multi:rq3}.

\subsubsection{Freezing rain and wind gust}
\label{sec:multi:fzg-method}

Unlike the nine primary hazards, freezing rain and concurrent wind gust constitute a compound hazard in which neither component alone determines network damage. Ice accumulation imposes static loading on conductors and support structures. At the same time, concurrent gusts amplify the dynamic load through wind-induced sway and torque. Damage thresholds and restoration durations scale jointly with both intensities \cite{coburn_2024_quantifying}. We characterize this compound hazard using the Sperry-Piltz Ice Accumulation (SPIA) index, a six-tier classifier (0--5). It maps the joint distribution of radial ice accumulation ($R_{eq}$) and concurrent wind gust ($G$) to discrete utility-impact descriptions. These descriptions range from minimal risk (tier 0) to catastrophic damage with multi-week outages (tier 5). The per-state 50-year return period values of $R_{eq,50}$, concurrent $G_{50}$, and the resulting $\textrm{SPIA}_{50}$ are obtained from \citeA{coburn_2024_quantifying}. They derived these values from 18 years (2005--2022) of hourly observations at 883 Automated Surface Observing System (ASOS) stations across CONUS. Hourly ice accumulation was computed using the freezing fraction implementation of the Cold Regions Research and Engineering Laboratory (CRREL) energy-balance model. This was combined with concurrent gust observations to produce hourly SPIA values. The 50-year return period quantiles were derived via Gumbel fits to annual maxima pooled across ASOS stations within each state. As with the geomagnetic layer, we adopt their published state-level outputs directly rather than reprocessing the underlying meteorological record. Each node and substation inherits the $R_{eq,50}$, $G_{50}$, and $\textrm{SPIA}_{50}$ of the state in which it lies via a spatial within-join against the US Census 2020 state polygons. The resulting FZG exposure of the network is shown in Figure~\ref{fig:multihazard_exposure}h.

\subsection{Fragility functions}
\label{sec:multi:fragility}

Fragility functions characterize the conditional probability that an asset reaches or exceeds a given damage state (DS) as a function of hazard intensity measure (IM) \cite{saouma_fragility_2021}. This is expressed with a lognormal formulation as
\begin{equation}
    P(DS \geq ds_i \mid IM = h) = \Phi\!\left[\frac{\ln(h/\theta_i)}{\beta_i}\right],
    \label{eq:fragility}
\end{equation}
\noindent
where $h$ is a specific value of IM, $ds_i$ is the threshold for damage state $i$, $\theta_i$ is the median capacity, $\beta_i$ is the logarithmic standard deviation, and $\Phi(\cdot)$ is the standard normal cumulative distribution function (CDF).

For earthquake, flood, and TC wind hazards, we adopt multi-state lognormal functions using a four-tier damage state classification: Slight, Moderate, Extensive, and Complete. This classification is consistent with the HAZUS methodology \cite{fema_hazus_earthquake_2022}. Threshold parameters $\theta$ and $\beta$ are drawn from the fragility function compilation of \citeA{kabre_fragility_2022}. This compilation includes earthquake substation parameters based on HAZUS anchored equipment classes by voltage tier. Flood parameters are adapted from \citeA{sanchez-munoz_electrical_2020} and \citeA{karagiannis_climate_2017}. Wind parameters come from HAZUS internal tables as reported by \citeA{watson_modeling_2020}, together with transmission line models from \citeA{huang_resilience-constrained_2018}. Damage ratios of 5\%, 20\%, 50\%, and 100\% are adopted as simplified values consistent with the HAZUS best-estimate ranges for electric power substations \cite{fema_hazus_earthquake_2022}. The HAZUS ranges are 0.01--0.15, 0.15--0.40, 0.40--0.80, and 0.80--1.00 for Slight through Complete states, respectively.

Standardized, published, multi-state fragility functions for high-voltage transmission lines and substations against tornado, wildfire, hail, lightning, and landslide hazards were not identified in the reviewed literature. For lightning and wildfire in particular, this gap is corroborated by \citeA{karagiannakis_fragility_2025}, who identify these as priority areas for further fragility research. \citeA{kabre_fragility_2022} similarly documents the absence of calibrated functions for these hazard types in the infrastructure resilience literature. We therefore adopt single-state scenario fragilities for these five hazards. Threshold parameters and damage ratios ($dr$) are selected on the basis of physical reasoning as follows.

\begin{itemize}
    \item \textbf{Tornado:} complete failure at a median EF-derived wind speed of 60--70\,m/s ($\beta = 0.3$), consistent with EF3+ winds that exceed the design wind loading of most transmission structures. The narrow $\beta$ reflects the well-defined relationship between tornado wind speed and structural failure.

    \item \textbf{Wildfire:} moderate damage at $\theta = 10{,}000$ WHP ($\beta = 0.5$, $dr = 0.30$) for substations, reflecting that substation equipment is largely non-combustible but vulnerable to radiant heat and ember exposure at very high hazard potential. Lines use a lower threshold ($\theta = 8{,}000$, $dr = 0.20$) given conductor and wood support structure exposure.

    \item \textbf{Hail:} slight damage at $\theta = 1.5$--$2.5$\,inches ($\beta = 0.4$, $dr = 0.03$--$0.05$), representing damage to insulators, bushings, and control equipment at diameters exceeding the severe hail threshold. Higher-voltage equipment is assumed to be more robust, leading to an increasing $\theta$ with increasing voltage tier.

    \item \textbf{Lightning:} moderate damage at $\theta = 10$ flashes/km$^2$/yr ($\beta = 0.5$, $dr = 0.05$--$0.10$), reflecting cumulative annual flashover and surge arrester degradation risk that scales with flash density. Lines are assigned a lower damage ratio ($dr = 0.05$) than substations ($dr = 0.10$) due to the higher concentration of equipment at substations.

    \item \textbf{Landslide:} complete failure at susceptibility index $\theta = 50$--$60$ ($\beta = 0.5$), representing foundation undermining or tower collapse in zones where multiple geomorphic susceptibility factors converge. Lines use a lower threshold ($\theta = 50$, $dr = 0.80$) than substations ($\theta = 60$, $dr = 1.00$), reflecting the distributed exposure of line corridors.
\end{itemize}

These scenario fragilities carry greater epistemic uncertainty than the HAZUS-derived functions. They should be interpreted as order-of-magnitude bounding estimates rather than calibrated damage predictions. The sensitivity of aggregate results to these parameters is discussed in Section~\ref{sec:multi:discussion}.

The FZG compound hazard uses a discrete-tier damage-state lookup rather than a continuous lognormal. The SPIA tiers $\{0, 1, 2, 3, 4, 5\}$ are used directly as the damage state classifier. Each tier carries a utility-grounded impact description \cite{coburn_2024_quantifying}. We map SPIA tiers to damage ratios as $\{0 \to 0.00,\ 1 \to 0.05,\ 2 \to 0.15,\ 3 \to 0.40,\ 4 \to 0.70,\ 5 \to 1.00\}$. This applies the same HAZUS damage-ratio convention used for the lognormal hazards. The failure threshold $\textrm{EDR} \geq 0.50$ corresponds to SPIA tier 4, defined as ``extensive damage to main feeder lines, some high-voltage transmission structures, outages lasting 5--10 days.'' This aligns with the same failure semantics used across all other hazards. Geomagnetic disturbance uses the per-substation failure probabilities of \citeA{oughton_major_2025}, derived from an asset-specific thermal fragility model and described in Section~\ref{sec:multi:hazards}.

The expected damage ratio (EDR), computed by weighting the discrete probability of occurrence for each damage state by its corresponding damage ratio \cite{kim_development_2024}, is
\begin{equation}
    \mathrm{EDR}(h) = \sum_{i=1}^{n} dr_i\bigl[P(DS \geq ds_i \mid h) - P(DS \geq ds_{i+1} \mid h)\bigr],
    \label{eq:edr}
\end{equation}
\noindent
where $dr_i$ is the damage ratio for state $i$, $n$ is the total number of damage states, and $P(DS \geq ds_{n+1}) \equiv 0$ by convention.

\begin{figure}[H]
    \centering
    \includegraphics[width=\textwidth]{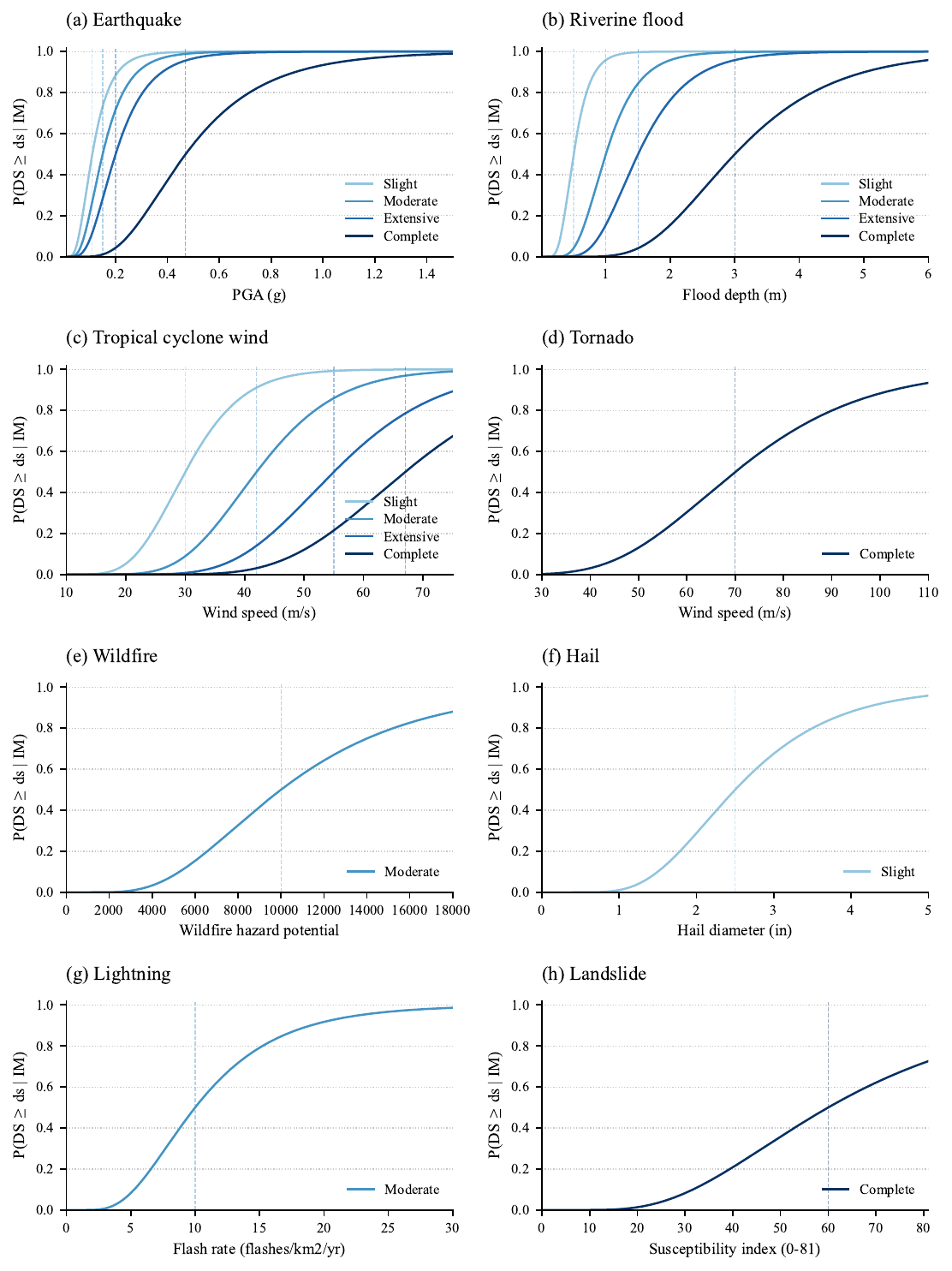}
    \caption{\textbf{Fragility curves for HV substations ($>$345\,kV) across the eight parametric-fragility hazards.} Panels (a)--(d) show multi-state lognormal curves from the literature synthesis. Panels (e)--(h) show single-state scenario curves based on physical reasoning. Dashed lines indicate median capacity ($\theta$). FZG (discrete tier lookup) and geomagnetic disturbance (asset-specific thermal fragility) are not parametric lognormals and are documented separately.}
    \label{fig:fragility}
\end{figure}

Substations are classified into three voltage tiers: low-voltage (LV, $\leq$161\,kV), medium-voltage (MV, 162--345\,kV), and high-voltage (HV, $>$345\,kV). Fragility parameters are adjusted across tiers following the same damage state structure. For brevity, Table~\ref{tab:fragility} presents representative parameters for HV substations only. Figure~\ref{fig:fragility} shows the corresponding fragility curves.

\begin{table}[H]
\centering
\caption{\textbf{Representative fragility function parameters for HV substations ($>$345\,kV, anchored).} $\theta$ is the median capacity, $\beta$ is the logarithmic standard deviation, and $dr$ is the damage ratio. Multi-state lognormal entries are taken from a literature synthesis \cite{kabre_fragility_2022}. Single-state scenario entries are derived from physical reasoning. The FZG entry is a discrete-tier lookup rather than a parametric fragility (Section~\ref{sec:multi:fragility}). Geomagnetic fragility is detailed in Section~\ref{sec:multi:hazards}.}
\label{tab:fragility}
\vspace{2mm}
\footnotesize
\setlength{\tabcolsep}{3pt}
\renewcommand{\arraystretch}{1.05}
\begin{tabular}{@{}l l l c c c p{3.2cm}@{}}
\hline
\textbf{Hazard} & \textbf{IM} & \textbf{Damage state} & $\theta$ & $\beta$ & $dr$ & \textbf{Basis} \\
\hline
\multirow{4}{*}{Earthquake}
  & \multirow{4}{*}{PGA (g)}
  & Slight    & 0.11 & 0.50 & 0.05 & \multirow{4}{=}{Literature synthesis \cite{kabre_fragility_2022}} \\
  & & Moderate   & 0.15 & 0.50 & 0.20 & \\
  & & Extensive  & 0.20 & 0.50 & 0.50 & \\
  & & Complete   & 0.47 & 0.50 & 1.00 & \\
\hline
\multirow{4}{*}{Flood}
  & \multirow{4}{*}{Depth (m)}
  & Slight    & 0.50 & 0.40 & 0.05 & \multirow{4}{=}{Literature synthesis \cite{kabre_fragility_2022}} \\
  & & Moderate   & 1.00 & 0.40 & 0.20 & \\
  & & Extensive  & 1.50 & 0.40 & 0.50 & \\
  & & Complete   & 3.00 & 0.40 & 1.00 & \\
\hline
\multirow{4}{*}{TC wind}
  & \multirow{4}{*}{Speed (m/s)}
  & Slight    & 30 & 0.25 & 0.05 & \multirow{4}{=}{Literature synthesis \cite{kabre_fragility_2022}} \\
  & & Moderate   & 42 & 0.25 & 0.20 & \\
  & & Extensive  & 55 & 0.25 & 0.50 & \\
  & & Complete   & 67 & 0.25 & 1.00 & \\
\hline
Tornado   & EF wind (m/s)           & Complete & 70    & 0.30       & 1.00      & Physical reasoning \\
Wildfire  & WHP (0--18000)          & Moderate & 10000 & 0.50       & 0.30      & Physical reasoning \\
Hail      & Diameter (in)           & Slight   & 2.50  & 0.40       & 0.05      & Physical reasoning \\
Lightning & Flash rate              & Moderate & 10    & 0.50       & 0.10      & Physical reasoning \\
Landslide & Susceptibility          & Complete & 60    & 0.50       & 1.00      & Physical reasoning \\
\hline
FZG       & SPIA (0--5)             & Discrete & --    & --         & 0.0--1.0  & SPIA tier lookup \cite{coburn_2024_quantifying} \\
Geomag    & GIC (A/phase)           & Failure  & 75    & 0.25--0.50 & --        & \citeA{oughton_major_2025} \\
\hline
\end{tabular}
\end{table}

\subsection{Asset inventory and replacement costs}
\label{sec:multi:inventory}

Replacement costs are estimated using unit costs from the Midcontinent Independent System Operator (MISO) transmission cost estimation guide for MTEP24 \cite{midcontinent_independent_system_operator_miso_transmission_2024}, expressed in 2023 US dollars. These values represent indicative planning-level estimates intended for regional economic evaluation rather than final construction bids. They carry inherent epistemic uncertainty under varying terrain and routing conditions. Each line's nominal voltage is snapped to the nearest standard level $v \in \{69, 115, 161, 230, 345, 500, 765\}$\,kV. Substation costs are assigned based on the maximum voltage of connected lines. Table~\ref{tab:miso_costs} summarizes the adopted unit costs.

\begin{table}[H]
\centering
\caption{\textbf{Unit replacement costs by voltage class} from MISO MTEP24 \cite{midcontinent_independent_system_operator_miso_transmission_2024}. All values represent indicative planning estimates in 2023 US dollars.}
\label{tab:miso_costs}
\vspace{2mm}
\footnotesize
\setlength{\tabcolsep}{5pt}
\renewcommand{\arraystretch}{1.0}
\begin{tabular*}{0.85\linewidth}{@{}c @{\extracolsep{\fill}} cc@{}}
\hline
\textbf{Voltage (kV)} & \textbf{Line cost (\$/mile)} &
\textbf{Substation cost (\$M)} \\
\hline
69  & 1{,}500{,}000 & 8.0  \\
115 & 2{,}500{,}000 & 15.0 \\
161 & 2{,}750{,}000 & 20.0 \\
230 & 3{,}000{,}000 & 37.5 \\
345 & 3{,}050{,}000 & 75.0 \\
500 & 3{,}600{,}000 & 150.0 \\
765 & 5{,}900{,}000 & 300.0 \\
\hline
\end{tabular*}
\end{table}

\subsection{Loss estimation}
\label{sec:multi:loss}

The expected annual loss (EAL) for each asset-hazard pair is
\begin{equation}
    \mathrm{EAL}_j = \mathrm{EDR}(h_j) \times C_j \times \lambda_j,
    \label{eq:eal}
\end{equation}
\noindent
where $j$ indexes each asset, $C_j$ is the asset replacement cost, $\mathrm{EDR}(h_j)$ is the expected damage ratio computed from Equation~\ref{eq:edr} at the asset's hazard intensity $h_j$, and $\lambda_j$ is the annual hazard occurrence rate at the asset's location. For hazards expressed as continuous intensity measures at a single return-period scenario, $\lambda_j = 1$. These hazards are earthquake, flood, landslide, wildfire, lightning, TC wind, geomagnetic disturbance, and FZG. For these return-period scenarios, the reported EAL should be interpreted as scenario-conditioned damage rather than exceedance-probability-weighted annualized loss. For event-based hazards with an empirical historical rate, $\lambda_j$ is the annual event rate at the asset's location. These hazards are tornado and hail, and their rates are derived from the storm-report record described in Section~\ref{sec:multi:hazards}. The expected daily damage (EAD) follows as $\mathrm{EAD}_j = \mathrm{EAL}_j / 365$.

All ten hazards are carried through to loss estimation. Assets with $\mathrm{EDR} \geq 0.50$ are classified as failed. Substations connected to failed transmission lines are additionally flagged as indirectly affected for the economic propagation step described in Section~\ref{sec:multi:econ}.

\subsection{Socioeconomic impact assessment}
\label{sec:multi:econ}

To quantify the societal impacts of power transmission failures, we adopt the spatial economic framework of \citeA{oughton_major_2025} and extend it to multi-hazard assessment. Substation service areas are defined via Voronoi tessellation of substation coordinates. Population and sectoral economic activity are spatially redistributed from ZIP Code Tabulation Areas (ZCTAs) to service areas using masked dasymetric interpolation weighted by the National Land Cover Database (NLCD) 2023 developed land cover classes \cite{usgs_national_2020}. We reuse the pre-computed interpolation tiles produced by \citeA{oughton_major_2025}. These provide per-substation population from the 2020 US Census \cite{bureau_2020_2020} and GDP across ten sectors aligned with the North American Industry Classification System (NAICS). The underlying sectoral GDP figures are taken from 2023 Bureau of Economic Analysis (BEA) state-level estimates \cite{bea_bea_2023}. They are disaggregated to the substation level by establishment density from the Statistics of US Businesses survey \cite{bureau_statistics_2023}.

For each hazard, the set of affected substations $F$ comprises those classified as failed either directly or through connection to a failed transmission line. The failure threshold is expected damage ratio $\mathrm{EDR} \geq 0.50$. The total affected population is
\begin{equation}
    L_{\mathrm{pop}} = \sum_{i \in F} P_i,
    \label{eq:pop}
\end{equation}
\noindent
where $P_i$ is the resident population within service area $i$.

Economic loss is propagated through a Leontief demand-driven input--output model, the appropriate formulation for tracing backward propagation of a final-demand cut through inter-industry purchases \cite{miller_input-output_2009}. The daily final-demand shock is a population-share scaling of household and government consumption,
\begin{equation}
    \Delta \mathbf{f} = -\frac{\rho}{365}\,\mathbf{f}_{\mathrm{cons}},
    \label{eq:demand_shock}
\end{equation}
\noindent
where $\mathbf{f}_{\mathrm{cons}} \in \mathbb{R}^{10}$ is the annual household plus government final-demand vector across the ten NAICS sectors (USD per year), $\rho = L_{\mathrm{pop}} / P_{\mathrm{grid}}$ is the fraction of the grid-served population affected, and $P_{\mathrm{grid}}$ is the total population served by the modeled transmission network. Pre-multiplying this shock by the Leontief inverse gives the vector of daily output losses across all sectors,
\begin{equation}
    \Delta \mathbf{x} = \mathbf{L}\,\Delta \mathbf{f}, \qquad \mathbf{L} = (\mathbf{I} - \mathbf{A})^{-1},
    \label{eq:leontief}
\end{equation}
\noindent
where $\mathbf{A} \in \mathbb{R}^{10 \times 10}$ is the direct-requirements matrix derived from the BEA 10-sector input--output accounts, $\mathbf{I}$ is the $10 \times 10$ identity matrix, and $\Delta \mathbf{x} \in \mathbb{R}^{10}$ is the resulting vector of daily output losses across the ten sectors. We report the total daily output loss $\sum_s \Delta x_s$, the per-sector direct and total losses $\Delta f_s$ and $\Delta x_s$, and the implied Leontief multiplier $m = \sum_s \Delta x_s / \sum_s \Delta f_s$.

The population-share allocation is a conservative simplification. It assumes uniform per-capita economic exposure within each affected service area. This understates concentration in dense urban centers, but it avoids the over-counting implicit in establishment-weighted allocation. Under establishment-weighted allocation, every business in an affected area is implicitly assumed to lose its full daily output. The static input--output structure further omits substitution, inventory buffering, and demand response. These factors would reduce realized losses during short outages. The resulting economic estimates are therefore upper bounds on daily losses, assuming no adaptive response.

\newpage
\section{Results}
\label{sec:multi:results}

\subsection{Hazard risk to the transmission network}
\label{sec:multi:rq1}

Across the ten hazards considered, the aggregate EAD to the US bulk transmission network is \$1,478\,M/day. Hazards are grouped by their temporal basis, and return-period (RP) scenarios represent losses at specified exceedance probabilities. These include earthquake at the 475-year RP, flood and TC wind at the 100-year RP, FZG at the 50-year RP, and geomagnetic disturbance at the 100-year RP.

\begin{figure}[H]
    \centering
    \includegraphics[width=\textwidth]{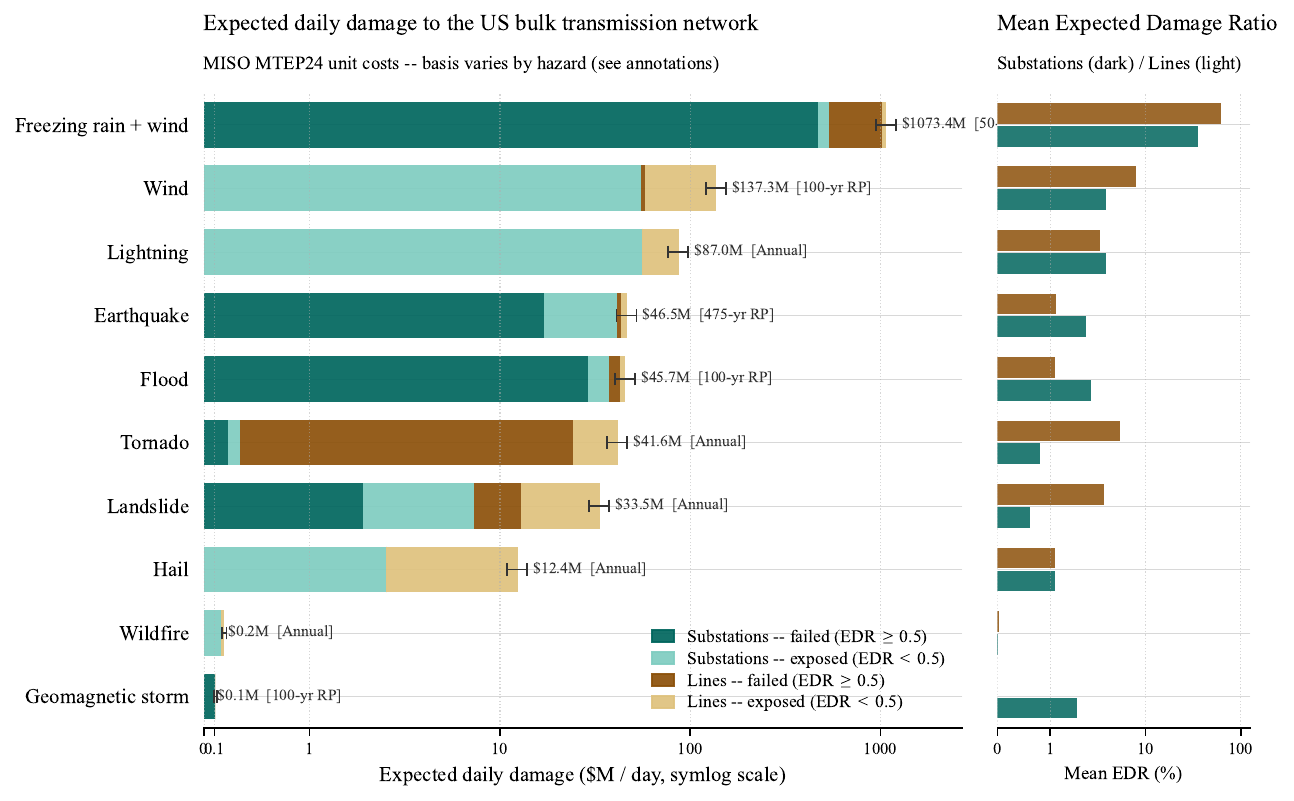}
    \caption{\textbf{Expected daily damage to the US bulk transmission network by hazard.} Left: stacked EAD decomposed by asset type (substations versus lines) and failure mode. Right: mean EDR by asset type.}
    \label{fig:ead}
\end{figure}

Figure~\ref{fig:ead} presents EAD ranked by magnitude. FZG dominates at \$1,073\,M/day, more than an order of magnitude above the next-largest hazard. This reflects two compounding factors. First, FZG is the only hazard for which intensity data are published at state-level spatial resolution. Every asset within a high-SPIA state, therefore, inherits the same tier. Second, SPIA tier 5 is reached across most of the Midwest and Northeast at the 50-year RP. A uniform damage ratio of 1.00 is then applied to thousands of assets.

The next tier is led by TC wind at \$137\,M/day, followed by lightning at \$87\,M/day, earthquake at \$47\,M/day, flood at \$46\,M/day, tornado at \$42\,M/day, and landslide at \$34\,M/day. TC wind reflects a 100-year RP scenario, concentrated in the tropical cyclone risk zone and covering 995 of 3,108 CONUS counties. Lightning and landslide are continuous annual exposures. Earthquake and flood are RP scenarios at 475-year and 100-year levels. Hail at \$12\,M/day, wildfire at \$0.2\,M/day, and geomagnetic disturbance at \$0.1\,M/day round out the lower end. The geomagnetic EAD reflects only the direct replacement cost of transformer-level failures.

The composition of damage varies markedly across hazards. Lightning, TC wind, hail, and wildfire predominantly cause sub-threshold damage. Their EDR values remain below 0.50, and few or no individual assets reach the failure classification despite widespread low-level exposure. In contrast, earthquake, flood, landslide, tornado, FZG, and geomagnetic disturbance generate failed assets. FZG produces by far the largest number of failures, with 7,280 affected substations and 484 failed lines. This is consistent with its outlier EAD. Tornado produces the largest number of line failures among the terrestrial hazards, with 24 failed lines and line-dominated EAD of \$24\,M/day. Flood produces the most failed substations relative to its EAD, with 342 substations affected. Landslide risk is predominantly borne by transmission lines, with a mean line EDR of 3.66\% versus 0.62\% for substations. This reflects the alignment of mountain corridors with high-susceptibility terrain. FZG exhibits the highest mean EDR for both substations at 35.16\% and lines at 62.81\%.

\begin{figure}[H]
    \centering
    \includegraphics[width=\textwidth]{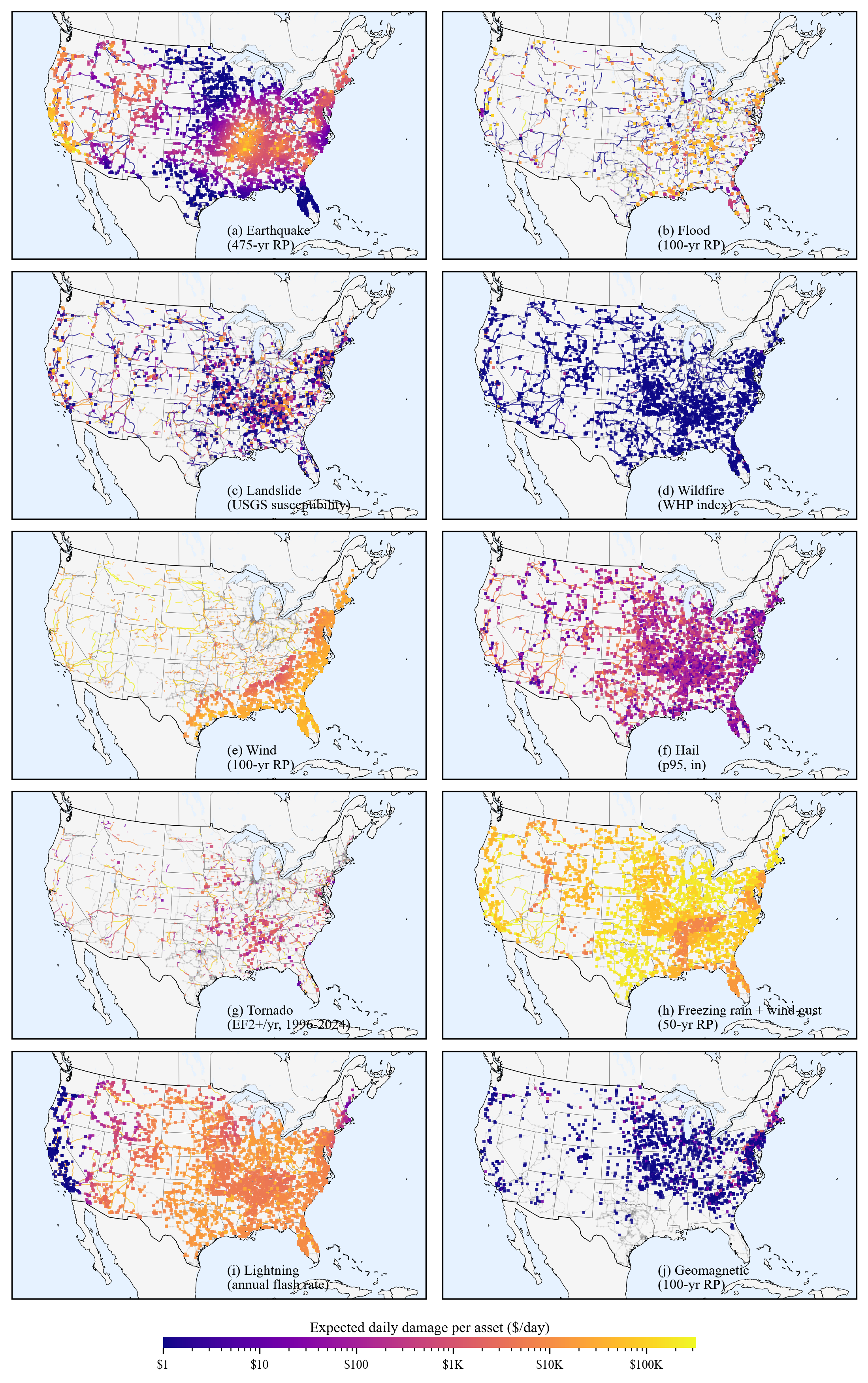}
    \caption{\textbf{Spatial distribution of expected daily damage per asset across ten hazards.} Substations (squares) and transmission lines (line segments) are colored by per-asset EAD on a shared log-normalized scale.}
    \label{fig:spatial_ead}
\end{figure}

Figure~\ref{fig:spatial_ead} maps per-asset EAD across CONUS. Earthquake damage is concentrated along the Pacific coast and in the New Madrid seismic zone. Flood damage follows major river systems, particularly the Mississippi and Ohio valleys. Landslide exposure spans the Appalachian and Rocky Mountain corridors. This produces high peak-per-asset losses because of steep susceptibility gradients along mountain transmission routes. Wildfire exposure is geographically extensive across the western US but generates low per-asset damage.

Lightning and TC wind exhibit the broadest spatial footprints among the terrestrial hazards. Lightning covers nearly all substations east of the Rockies, with peak intensity in the Southeast. TC wind damage is concentrated along the Gulf Coast and Atlantic seaboard. Hail exposure tracks the Great Plains corridor from Texas through Nebraska. Tornado produces the most spatially concentrated damage, with the highest per-asset magnitude among the terrestrial hazards. This is driven by direct EF2+ track intersections with transmission corridors across the tornado-prone region spanning the Deep South and the central Great Plains. FZG saturates the eastern half of CONUS at the failure tier, with hotspot assets covering the Midwest and Northeast. Geomagnetic exposure concentrates along the Atlantic seaboard and in regions of high ground conductivity contrast. This pattern is consistent with the geoelectric-field pattern.

\subsection{Economic loss propagation}
\label{sec:multi:rq2}

\begin{figure}[H]
    \centering
    \includegraphics[width=\textwidth]{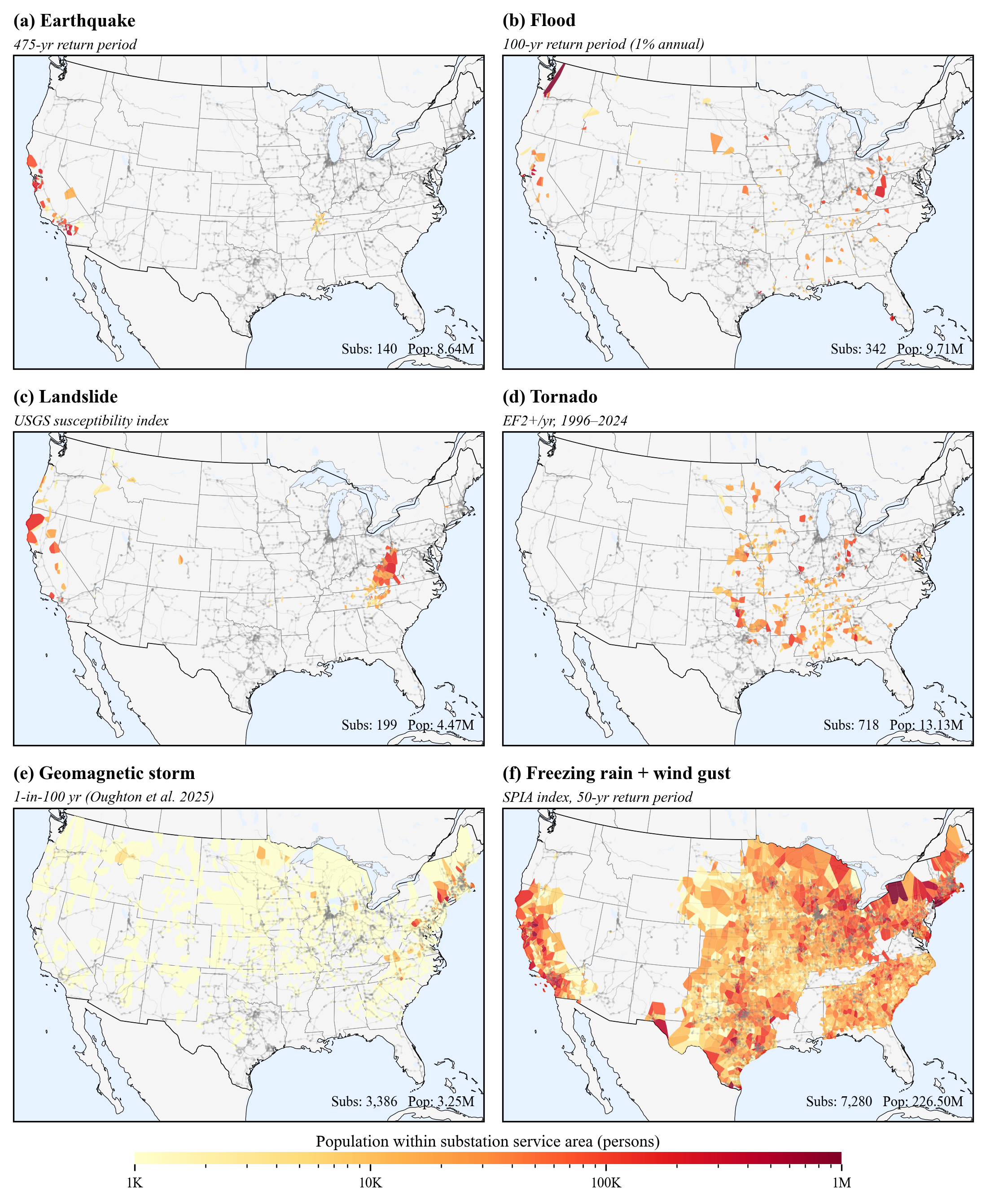}
    \caption{\textbf{Population exposure within affected substation service areas.} Each polygon is a substation Voronoi service area, colored by the resident population it serves on a log scale from 1K to 1M persons.}
    \label{fig:pop}
\end{figure}

Figure~\ref{fig:pop} maps the population within affected substation service areas across the hazards producing asset failures. FZG affects 237.4\,M people across 7,280 substations, reflecting the state-level granularity of the hazard layer. Tornado affects 13.75\,M across 718 substations. Flood affects 10.01\,M across 342 substations, earthquake affects 8.43\,M across 140, geomagnetic disturbance at the 250-year RP affects 5.78\,M across 3,909 substations, landslide affects 4.94\,M across 199, geomagnetic disturbance at the 100-year RP affects 4.12\,M across 3,386 substations, and TC wind affects 4.52\,M across 84. Earthquake and flood affect comparable populations to tornado despite having fewer failed substations because their failures concentrate in densely populated metropolitan areas. Geomagnetic disturbance affects many substations across a broad area, but the per-substation population is smaller on average. The affected pattern includes less densely populated regions.

\begin{figure}[H]
    \centering
    \includegraphics[width=\textwidth]{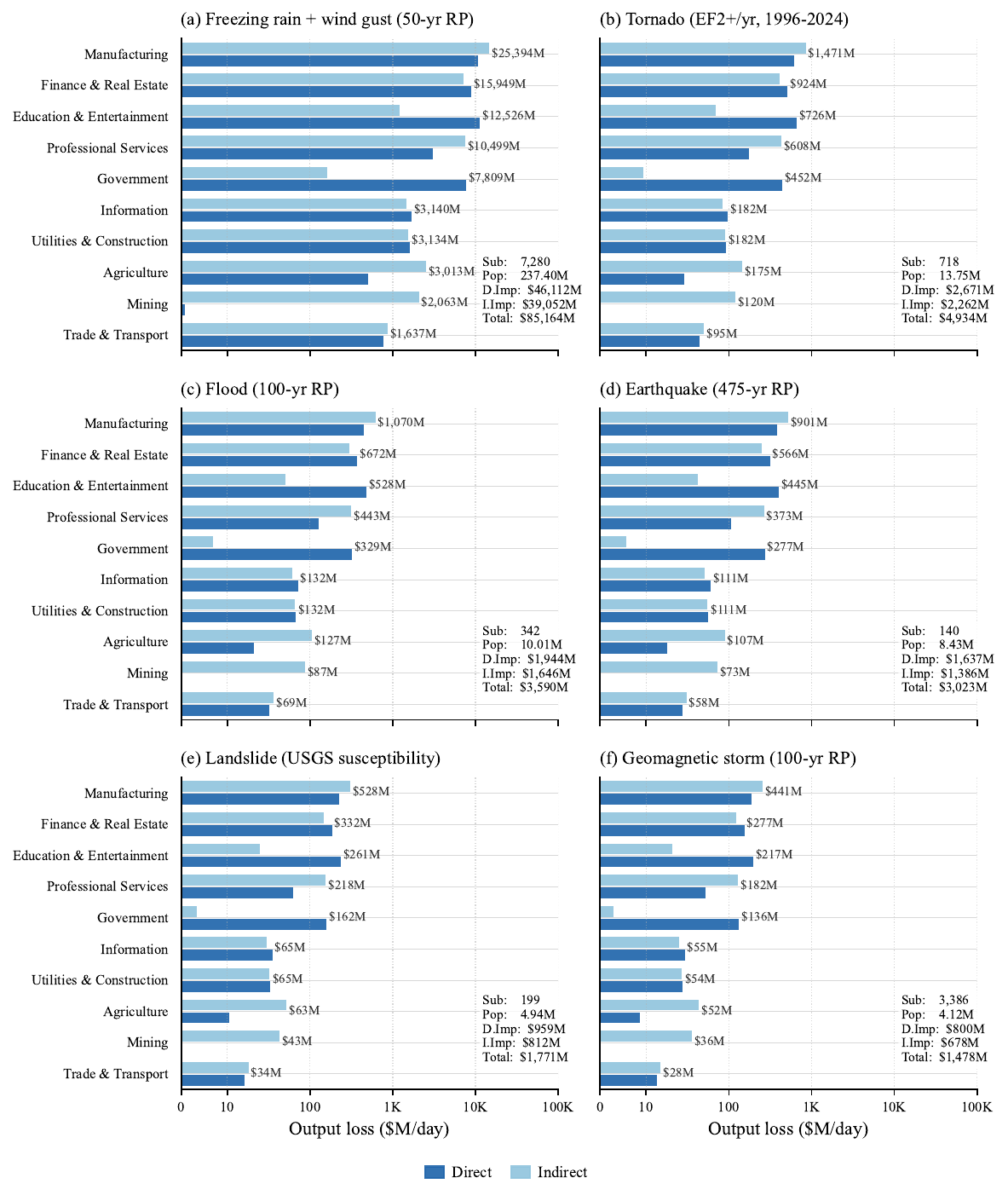}
    \caption{\textbf{Sectoral economic output loss from transmission network disruption.} Side-by-side bars show direct final-demand shock (dark) and indirect input--output cascade loss (light) by NAICS sector on a symlog x-axis. \textbf{Sub} indicates affected substations, \textbf{Pop} indicates affected population, and \textbf{D.Imp} and \textbf{I.Imp} denotes direct and indirect impacts, respectively.}
    \label{fig:io}
\end{figure}

Figure~\ref{fig:io} presents sectoral economic losses decomposed into direct final-demand shock and indirect input--output cascade effects. FZG produces \$85,164\,M/day in total downstream output losses, with \$46,112\,M direct and \$39,052\,M indirect. The largest sectoral impacts are manufacturing at \$25,394\,M, finance and real estate at \$15,949\,M, education and entertainment at \$12,526\,M, professional services at \$10,499\,M, and government at \$7,809\,M. This figure is an upper bound consistent with the state-level granularity of the FZG hazard layer. Tornado produces \$4,934\,M/day in total losses, with manufacturing at \$1,471\,M and finance and real estate at \$924\,M leading sectoral impacts. Flood produces \$3,590\,M/day, with manufacturing at \$1,070\,M and finance and real estate at \$672\,M. Earthquake produces \$3,023\,M/day, with manufacturing at \$901\,M and finance and real estate at \$566\,M. Geomagnetic disturbance at the 250-year RP produces \$2,074\,M/day, placing it between flood and landslide in economic magnitude. Landslide produces \$1,771\,M/day, with manufacturing at \$528\,M leading. TC wind produces \$1,622\,M/day. Geomagnetic disturbance at the 100-year RP produces \$1,478\,M/day, with manufacturing at \$441\,M and finance and real estate at \$277\,M as the most affected sectors.

Manufacturing, finance and real estate, education and entertainment, professional services, and government emerge as the five most affected sectors across all hazards considered. The sectoral ranking is stable because the final-demand shock is allocated by population share. The underlying household and government consumption vector also has a consistent sectoral composition.

The Leontief multiplier is 1.85$\times$ across all hazards. Each dollar of direct final-demand shock generates an additional \$0.85 in cascading losses through inter-sectoral demand linkages. The multiplier stability reflects the structure of the BEA 10-sector input--output accounts. The economic amplification depends on the sectoral composition of the final-demand shock, not on the geographic pattern of failures.

\subsection{Space weather versus terrestrial hazards}
\label{sec:multi:rq3}

A 100-year geomagnetic storm produces \$1,478\,M/day in economic losses, and a 250-year storm produces \$2,074\,M/day. The 250-year scenario ranks below earthquake, flood, and tornado in terms of daily economic impact. It ranks above landslide and TC wind. Population exposure at the 250-year level is 5.78\,M, lower than the corresponding terrestrial hazards. The terrestrial hazards represent annual or RP baselines, while the 250-year geomagnetic storm is a much rarer event. On a risk-adjusted basis, the expected annual impact of space weather is therefore lower than the leading terrestrial hazards. The losses from a single extreme event nonetheless approach those of a 100-year flood or 475-year earthquake.

\newpage
\section{Discussion}
\label{sec:multi:discussion}

\subsection{Hazard risk to the transmission network}
\label{sec:multi:disc_rq1}

FZG dominates EAD at \$1,073\,M/day, more than an order of magnitude above any other hazard. Among the terrestrial hazards with finer spatial resolution, TC wind leads at \$137\,M/day, followed by lightning at \$87\,M/day. Lightning is considered a continuous annual exposure. This is consistent with the empirical finding of \citeA{rimkus_impact_2024} that it is a substantial contributor to transmission outages alongside wind and precipitation. Earthquake at \$47\,M/day and flood at \$46\,M/day fall in a comparable band. Tornado at \$42\,M/day ranks below the continuous-exposure layers, despite producing the largest line-failure count among the terrestrial hazards.

Flood causes the most failures among any single-RP hazard, with 342 substations affected. This reflects concentrated riverine exposure. Landslide risk is predominantly line-borne, with a mean line EDR of 3.66\% versus 0.62\% for substations. This is consistent with the vulnerability of transmission corridors crossing landslide-prone terrain documented by \citeA{ghorani_modeling_2021}. TC wind, lightning, hail, and wildfire generate predominantly sub-threshold damage. They produce widespread low-level degradation rather than discrete failures. This pattern aligns with the review by \citeA{hou_review_2022}, which notes that distribution- and transmission-level vulnerabilities tend to manifest cumulatively rather than as discrete failure events at extreme intensities.

Published multi-state fragility curves for tornado, wildfire, hail, lightning, and landslide applied to high-voltage infrastructure were not identified in the literature. This gap is highlighted by \citeA{karagiannakis_fragility_2025}. The wildfire EAD of \$0.2\,M/day appears low relative to observed impacts on western US grids reported by \citeA{vahedi_wildfire_2025}. The tornado fragility threshold of 70\,m/s also excludes damage from EF1 and EF2 events that \citeA{braik_risk-based_2019} showed can compromise aging poles.

\subsection{Economic loss propagation}
\label{sec:multi:disc_rq2}

FZG generates the largest total economic impact at \$85,164\,M/day. This is an upper bound consistent with the state-level granularity of the hazard layer. Among the terrestrial hazards with finer spatial resolution, tornado leads at \$4,934\,M/day, followed by flood at \$3,590\,M/day and earthquake at \$3,023\,M/day. Geomagnetic disturbance at the 250-year RP produces \$2,074\,M/day, placing it between flood and landslide. The 100-year scenario produces \$1,478\,M/day. Landslide at \$1,771\,M/day and TC wind at \$1,622\,M/day complete the comparable middle tier.

The Leontief multiplier is stable across all hazards at 1.85$\times$. This is consistent with the finding of \citeA{oughton_quantifying_2017} that indirect supply chain effects account for approximately half of total macroeconomic losses from transmission failure. The multiplier stability reflects the structure of the BEA 10-sector input--output accounts, not the geographic pattern of failures.

Manufacturing, finance and real estate, education and entertainment, professional services, and government emerge as the five most affected sectors across all hazards. This ranking is stable because the final-demand shock is allocated by population share. The underlying household and government consumption vector also has a consistent sectoral composition. The same population-share allocation that enables cross-hazard comparison, therefore, also produces a uniform sectoral signature. Differences between hazards manifest in the magnitude of the shock, not its sectoral distribution.

\subsection{Space weather versus terrestrial hazards}
\label{sec:multi:disc_rq3}

The 250-year geomagnetic storm ranks below earthquake, flood, and tornado in daily economic terms. It ranks above landslide and TC wind. The terrestrial estimates represent annual or return-period baselines, while the 250-year geomagnetic storm is a considerably rarer event. The losses from a single extreme event nonetheless approach those of a 100-year flood or 475-year earthquake.

\newpage
\section{Conclusion}
\label{sec:multi:conclusion}

This study applied a consistent hazard-exposure-fragility-economic framework across nine primary natural hazards and one compound hazard on the US high-voltage transmission network. The framework uses a common network representation, fragility database, and Leontief input--output economic model throughout. The aggregate EAD across the ten hazards is \$1,478\,M/day, dominated by FZG at \$1,073\,M/day at the 50-year RP. TC wind, lightning, and earthquake lead the next tier at \$137\,M, \$87\,M, and \$47\,M/day. Downstream economic losses are largest for FZG at \$85,164\,M/day and for the terrestrial RP hazards at \$4,934\,M/day for tornado, \$3,590\,M/day for flood, and \$3,023\,M/day for earthquake. A 250-year geomagnetic storm produces \$2,074\,M/day, placing space weather alongside the leading terrestrial threats at extreme return periods. The Leontief multiplier of 1.85$\times$ across all hazards demonstrates that indirect demand-linkage losses nearly match direct disruption costs.

These findings come with important caveats. First, the spatial granularity of input hazard layers varies by an order of magnitude across the ten hazards, from 1\,km flood depth to state-level FZG. The FZG result, in particular, is best interpreted as an upper bound rather than a calibrated point estimate because every asset within a high-SPIA state inherits the same damage tier regardless of local exposure. Second, the population-share allocation of the final-demand shock produces a stable sectoral ranking across all hazards. This is conservative because it avoids the over-counting implicit in establishment-weighted allocation, but it does not capture genuine regional variation in industrial mix. Third, calibrated multi-state fragility functions for tornado, wildfire, hail, lightning, and landslide applied to high-voltage infrastructure were not identified in the literature. The scenario fragilities adopted for these hazards rest on physical reasoning rather than empirical calibration. The wildfire and tornado results in particular likely understate the true impact, given the EF1 and EF2 vulnerabilities documented by \citeA{braik_risk-based_2019} and the western US fire-grid impacts reported by \citeA{vahedi_wildfire_2025}. Fourth, the framework does not account for restoration duration or recovery costs. The true economic burden of prolonged outages is therefore likely understated, particularly for flood and earthquake where substation repair can take weeks. Fifth, hazards are evaluated independently, which likely understates damage where compound events interact, as discussed by \citeA{macheri_assessing_2025} and \citeA{jackson_spatio-temporal_2025}. Sixth, the static input--output model omits adaptive responses, such as substitution, inventory buffering, and demand response. These would reduce realized losses during short outage durations.

This study demonstrates that a meaningful cross-hazard comparison of transmission network risk is both feasible and policy-relevant. By applying a single consistent framework from physical hazard to economic consequence, it provides infrastructure operators and policymakers a common currency for comparing threats that have, until now, been assessed in isolation. The absolute magnitudes reported here should be interpreted as bounding estimates. The cross-hazard ranking and the relative position of space weather are robust to the underlying assumptions.

\acknowledgments
This material is based upon work supported by the NSF National Center for Atmospheric Research, which is a major facility sponsored by the US National Science Foundation under Cooperative Agreement No. $1852977$. The project upon which this article is based was funded through the NSF NCAR Early-Career Faculty Innovator Program under the same Cooperative Agreement. We also gratefully acknowledge funding from the ChronoStorm NSF RAPID grant (\#$2434136$), co-funded by the GEO/AGS Space Weather Research and the ENG/CMMI Humans, Disasters, and the Built Environment programs. Additional support was provided through a supplemental award from the NSF NCAR Faculty Innovator Program covering October 2025--September 2026.

\section*{Open Research}

The analysis code, processing scripts, and visualization pipeline supporting this study are openly available in the GitHub repository (\url{https://github.com/denniesbor/mhtran}). The processed and derived datasets required to reproduce the main results and figures are archived at Zenodo \cite{bor_2026_20331026}.

The geomagnetic hazard data, transmission network topology, and input--output economic tiles are available in the coupled space weather impact (C-SWIM) GitHub repository (\url{https://github.com/denniesbor/C-SWIM}) and archived at Zenodo \cite{cswim_zenodo_2026}. Transmission line data are from the Homeland Infrastructure Foundation-Level Data transmission lines layer \cite{hifld_transmission_2023}, and substation and power-tower locations are from OpenStreetMap \cite{openstreetmap_openstreetmap_2024}. Earthquake hazard data are from the 2023 US Geological Survey National Seismic Hazard Model \cite{mark_d_petersen_data_2024}. Flood depth data are from the World Resources Institute Aqueduct Global Flood Model \cite{reig_aqueduct_2013}. Landslide susceptibility data are from the US Geological Survey national landslide assessment \cite{gina_m_belair_slope-relief_2024}. Wildfire hazard potential data are from the US Department of Agriculture Forest Service \cite{dillon_wildfire_2023}. Tropical cyclone wind data are from the STORM dataset \cite{bloemendaal_globally_2022}. Hail and tornado reports are from the NOAA Storm Prediction Center \cite{noaa_storm_prediction_center_noaa_2024-1,noaa_storm_prediction_center_noaa_2024}. Lightning climatology data are from the NASA LIS/OTD combined satellite climatology \cite{nasa_earth_science_data_systems_lisotd_2024}. Freezing rain and wind gust data are from the SPIA-based hazard atlas of \citeA{coburn_2024_quantifying} and its Zenodo archive \cite{coburn_2023_10080809}. Replacement costs are based on the MISO transmission cost estimation guide \cite{midcontinent_independent_system_operator_miso_transmission_2024}. Population, land-cover, and economic data are from the 2020 US Census \cite{bureau_2020_2020}, the National Land Cover Database \cite{usgs_national_2020}, Bureau of Economic Analysis estimates \cite{bea_bea_2023}, and Statistics of US Businesses data \cite{bureau_statistics_2023}.

\section*{Conflict of Interest disclosure}
The authors declare no competing interests.

\newpage
\bibliography{references}

\FloatBarrier
\end{document}